\documentclass[12pt,amssymb,amsmath,aps, english,showpacs]{revtex4}
\usepackage{subfigure}
\usepackage{graphicx}
\usepackage{epstopdf}
\usepackage[]{natbib}
\usepackage{epsfig}

\renewcommand{\b}[1]{\boldsymbol{#1}}

\newcommand{\grad}{\b{\nabla}}

\providecommand{\boldsymbol}[1]{\mbox{\boldmath $#1$}}

\usepackage{babel}

\begin{document}

\preprint{APS/123-DPP}

\title{Particle Energization in 3D Magnetic Reconnection of Relativistic Pair Plasmas}

\author{Wei Liu\footnote{Present address: Department of Radiation Physics, University of Texas M. D. Anderson Cancer Center, Houston, TX USA 77030;  wliu3@mdanderson.org}}

\affiliation{Los Alamos National Laboratory, Los Alamos, New Mexico, USA 87545}

\author{Hui Li\footnote{Also Center for Magnetic Self-Organization in Laboratory and Astrophysical Plasmas}}

\affiliation{Los Alamos National Laboratory, Los Alamos, New Mexico, USA 87545}

\author{Lin Yin}

\affiliation{Los Alamos National Laboratory, Los Alamos, New Mexico, USA 87545}

\author{B. J. Albright}

\affiliation{Los Alamos National Laboratory, Los Alamos, New Mexico, USA 87545}

\author{Edison P. Liang}

\affiliation{Department of Physics and Astronomy, Rice University, Houston, Texas, USA 77005}

\author{K. J. Bowers \footnote{Guest Scientist. Present address: D. E. Shaw Research, LLC, New York, NY, USA 10036}}

\affiliation{Los Alamos National Laboratory, Los Alamos, New Mexico, USA 87545}

\date{\today}

\begin{abstract}
We present large scale 3D particle-in-cell (PIC) simulations to examine particle energization in magnetic reconnection of relativistic electron-positron (pair) plasmas. The initial configuration is set up as a relativistic Harris equilibrium without a guide field. These simulations are large enough to accommodate a sufficient number of tearing and kink modes. Contrary to the non-relativistic limit, the linear tearing instability is faster than the linear kink instability, at least in our specific parameters. We find that the magnetic energy dissipation is first facilitated by the tearing instability and followed by the secondary kink instability. Particles are mostly energized inside the magnetic islands during the tearing stage due to the spatially varying electric fields produced by the outflows from reconnection. Secondary kink instability leads to additional particle acceleration. Accelerated particles are,
however,  observed to be thermalized quickly. 
The large amplitude of the vertical magnetic field resulting from the tearing modes by the secondary kink modes further help thermalizing the non-thermal particles generated from the secondary kink instability. Implications of these results for astrophysics are briefly discussed.  
 
\end{abstract}


\maketitle

\section{Introduction}

Magnetic reconnection in collisionless relativistic electron-position (pair) plasma plays an essential role in a number of important astrophysical problems, such as relativistic jets from supermassive black holes, relativistic winds from pulsars and possibly outflows from Gamma-ray Bursts. In the Crab pulsar wind, relativistic reconnection in pair plasmas has been considered as a primary mechanism to convert magnetic energy to particle energy. \cite{ks03,zh05} Although magnetic reconnection has been extensively studied, the physics of fast reconnection and particle energization in 3D relativistic reconnection is still a mystery. Magnetic reconnection of pair plasmas provides a unique opportunity to examine critically the fundamental reconnection physics due to the absence of Hall term, \cite{bb05} which is generally thought to be the main mechanism for fast reconnection in electron-ion plasmas. Understanding of relativistic magnetic reconnection of pair plasmas will also provide useful physical insights for both new laboratory experiments of pair plasmas \cite{cwb09} and observations of the high energy Universe. 

Previous reconnection studies of relativistic pair plasmas have focused on the fast reconnection and particle acceleration mostly in 2D \cite{zh01, zh05, bb07, zh07} or reconnection onset in 3D with relatively small system size. \cite{jtls04, zh05b, zh08}  It has been shown that instabilities in a cross-field plane such as the Kelvin-Helmholtz instability (KHI) \cite{yda96,ssf01} and the drift kink instability (DKI) \cite{dw98,zh05} are of critical importance to understanding magnetic energy dissipation. \citep{zh05} 
The past studies used small-scale particle-in-cell (PIC) simulations, (e.g., the length in the outflow direction is $\sim25$ inertial lengths), 
however, the characteristic system size in most astrophysical/laboratory systems is vastly larger. 
Thus, it is important to study these processes in a truly 3D configuration. 
Fast reconnection in the relativistic limit in 2D has been reported to be mostly facilitated by pair pressure tensors. \cite{hz07,bb07} \citet{ydk08} performed the largest 3D PIC simulations to date of non-relativistic pair plasma reconnection and found that elongated current sheet is unstable to secondary kink instability and leads to the bending of the current sheet. For particle acceleration, past 2D simulations of reconnection mostly started from an initially externally imposed reconnection electrical field and showed that particles are primarily accelerated near the X lines by electrical fields. \cite{hmts01,pp06,zh01,zh07} However, electrical fields within a small fraction of the total volume at X lines are difficult to account for the large number of accelerated electrons of the total volume. \cite{cbp08} Recently \citet{cbp08} reported some surprising satellite observation evidences that energetic electron flux peaks at sites of compressed density within islands during electon-ion reconnection in the Earth's magnetosphere. 
\citet{hz07} reported that for relativistic pair plasmas in 2D without a guide field, most of particle acceleration happens inside magnetic islands possibly like a betatron (but no details are provided) (also see \citet{dms04}). Thus, the well accepted particle acceleration mechanism needs to be revisited, especially in 3D. \citet{dscs06} suggests that Fermi acceleration from contracting magnetic islands is the major mechanism for acceleration in 3D non-relativistic electron-ion plasma magnetic reconnection and electrons are mostly accelerated within the islands.  
In this paper, we will present the first numerical study of 3D magnetic reconnection of relativistic pair plasma with an emphasis on particle energization. 

The paper is organized as follows. Section~\ref{setup} describes the
computational model and problem setup.  In Sec.~\ref{results},
overall simulation evolution is briefly described followed by particle energization and its mechanisms, particle thermalization and the role of the secondary kink instability on particle acceleration in magnetic reconnection. 
Conclusions and their implications for astrophysical situations are
given in Sec.~\ref{conclusion}.

\section{Simulation Set-up}\label{setup}

The geometry of the simulation is given in Fig. \ref{fig:geo}. 
The simulations are initialized with a relativistic Harris equilibrium current sheet. \cite{ks03,zh07} The initial magnetic field has
$B_y=B_0\tanh(x/L)$ and $B_x = B_z = 0$. Here $L$ is the half thickness of the current sheet. The initial plasmas consist of two 
distinct parts:
\begin{equation}
f_s(\mathbf{x},\mathbf{p})=\frac{n_0\cosh^{-2}(x/L)}{4\pi m_s^2cTK_2(m_sc^2/T)}\exp[-\frac{\gamma_s(\epsilon-\beta_sm_scu_z)}{T}]+\underline{\frac{n_{\rm b}}{4\pi m_s^2cT_bK_2(m_sc^2/T_b)}\exp(-\frac{\epsilon}{T_b})}\,,
\end{equation}
where the subscript $s$ denotes positron ($i$) and electron $e$, respectively. $u_z$ is the $z$-component of the four-speed, $\epsilon$ is the energy of the particle and $K_2(x)$ is the modified Bessel function of the second kind. One part (the underlined term) is a uniform background plasmas with density $n_b=0.3n_0$ and temperature $T_b=T_e=T_i=m_ec^2$ where $c$ 
is the speed of light and $m_e=m_i=m$ are the mass of electron and position, respectively. The second part of plasmas 
(for both electrons and positrons) has the same temperature $T=m_ec^2$ as the background plasma but their density is given as 
$n=n_0\cosh^{-2}(x/L)$. Electrons and positrons are drifting against each other, with electrons having $+V_d$ and 
positrons having $-V_d$. The drift velocity $V_d$ is given as $\beta=V_d/c= 0.82$. Both species have relativistic 
Maxwellian distributions. The initial current from the drifting distributions is in the $z$ direction. 
The simulation parameters are $L/d_i=0.7$ and $\omega_p/\Omega_c=0.5$, where $d_i=c/\omega_p$, 
$\omega_p=\sqrt{(4\pi n_0e^2)/m_i}$ and $\Omega_c=eB_0/(m_ic)$ are the positron inertial length, plasma frequency 
and cyclotron frequency, respectively. Periodic boundary conditions are used in $y$ and $z$ with reflecting boundaries in $x$. 

In such anti-parallel geometry, the linear Vlasov theory predicts two types of instabilities: tearing, 
with wave vectors along $y$, and kink, along $z$. \cite{dw99}
Using the parameters given above, linear theory predicts that the fastest growing tearing mode has wave number $k_yL=0.4$ and 
growth rate $\Gamma/\Omega_c=0.069$, while the fastest growing kink mode has $k_zL=0.25$ and $\Gamma/\Omega_c=0.059$. 
We have also performed 2D relativistic nonlinear tearing 
simulation (one cell in $z$ direction) and 2D relativistic nonlinear kink simulation (one cell in $y$ direction). 
We find that the fastest tearing mode has $k_yL=0.4$ and $\Gamma/\Omega_c=0.063$, and the fastest kink mode has $k_zL=0.25$ and $\Gamma/\Omega_c=0.053$. They match the linear Vlasov theory very well. 
\citet{pk07} also carried out similar linear Vlasov analysis in the relativistic limit (see also \cite{pk07b}). 
Compared to the results in the non-relativistic regime, \cite{dw99} the growth rates of both modes are greatly reduced and the wavelengths of both modes are increased by the relativistic effects, which is further verified in the 3D simulations. While in the non-relativistic regime the growth rate of the fastest growing kink mode is almost twice as large as that of the fastest growing tearing mode, the growth rate of the fastest tearing modes in the relativistic regime is a bit larger than the fastest growing kink modes, which results in the earlier emergence of tearing modes than kink modes in 3D simulations, contrary to that in the non-relativistic regime. 

In 3D, to examine how the reconnection dynamics may be influenced by both the linear and nonlinear competition between the tearing instability (in $x$-$y$ plane) and the kink instability (in $x$-$z$ plane), we use four simulations with the same size in the tearing plane $L_x\times L_y=200d_i\times200d_i$ but different sizes in the $z$ direction so that different spectra of kink modes can be excited. They include $L_z=0.195d_i$ (only 2D tearing is present), $20d_i$, $50d_i$, and $200d_i$, which are referred to as run A, B, C and D, respectively.  
These simulations follow the dynamics of $0.168$, $16$, $43$, and $84$ billion particles on $1024\times1024\times1$, $1024\times1024\times96$, $1024\times1024\times256$, and $1024\times1024\times500$ meshs, respectively. For these simulations, the cell sizes are: $\delta x=\delta y=0.195\lambda_D$, while $\delta z$ is $0.195\lambda_D$, $0.208\lambda_D$, $0.195\lambda_D$, and $0.4\lambda_D$, respectively. 
Here, $\lambda_D$ is the Debye length, and the time step $\delta t\Omega_c$ is $0.223$, $0.228$, $0.223$, and $0.258$, respectively. 
The total energy is conserved to better than $0.1\%$ throughout the simulation runs.  In order to remove the recirculation issue, all data presented here are chosen before recirculation happens ($\Omega_ct_{\rm recirculation}\sim\Omega_cL_y/v_{g,y}\sim4,000$, where $v_{g,y}\sim0.1$ is the typical particle group velocity in $y$-direction).

\section{Results}\label{results}

\subsection{Overall Evolution}

We now discuss the dynamics of the evolution. 

Fig.~\ref{fig:total} shows the time evolution of total magnetic energy of the four simulations. We have divided the magnetic energy conversion evolution into two main stages (separated by the dash line in Fig.~\ref{fig:total}).  For the first stage (tearing stage thereafter) when $t\Omega_c \lesssim 954$, the dynamics is dominated by the development of relativistic tearing instability as indicated by the red and green curves. The linear kink instability plays essentially little role for energy conversion in run A and B. It becomes visible in run C and D but it seems subdominant to the tearing modes (it has smaller growth rate compared to the tearing mode). The purely linear growth stage lasts until $t\Omega_c \sim 250$, then the islands from tearing start to merge, entering the nonlinear stage. The time $t\Omega_c \sim 954$ marks the eventual saturation of the tearing nonlinear evolution. 
For the second stage (secondary kink stage thereafter) when $t\Omega_c \gtrsim 954$, a new nonlinear instability becomes active. This was referred to as the secondary kink instability in an earlier, similar study using non-relativistic pair plasmas. \cite{ydk08} This happens in 3D only when there is sufficient length in the z-direction, as indicated by run C and D (black and blue curves). We can see that the tearing instability converts about 27\% of the initial magnetic energy into particle energy, followed by the additional conversion from $\sim27\%$ to $\sim40\%$ of the initial magnetic energy  by the secondary kink for run C and D. This result indicates that the overall energy conversion depends on how many modes are allowed in the z-direction and their nonlinear interactions with the tearing instability.

The relativistic cases presented here show a number of similarities to the previous non-relativistic 3D cases studied in \citet{ydk08}. During the linear phase $t\Omega_c \lesssim 250$ (Fig.~\ref{fig:tearkink1}), 
about $14$ small magnetic islands form along the $(x,y)$ plane from the tearing instability. Contrary to the non-relativistic limit, \cite{ydk08} the primary kink mode with wavelength $\sim 18.5d_i$ appears a little bit later than the tearing mode, undulating the small magnetic islands in the $z$ direction (Fig.~\ref{fig:tearkink1}). Then the magnetic islands formed from the tearing instability coalesce to produce a dominant reconnection site through which most of the magnetic flux is processed ($t\Omega_c \lesssim 954$) (Fig.~\ref{fig:tearkink2}).  When magnetic islands are formed,  the islands are not observed to be contracting, but rather they are either expanding or merging or remaining 
in a quasi-steady state. For run D, we see that, in the linear phase, both the tearing and kink instabilities are growing, causing the the reconnection sites to be "patchy" [Fig.~\ref{fig:tearkink2} (c)]. After the major reconnection site is established ($t\Omega_c \gtrsim 954$) (Fig.~\ref{fig:tearkink3}), this dominant diffusion region expands in $y$ to form a highly elongated current sheet unstable to secondary kink instability if the dimension in the current direction ($z$) is large enough. Note that the wavelength $\sim40d_i$ of the secondary kink mode is larger than the wavelength of the linear kink mode, thus there is no secondary kink modes in run B  $(L_z=20d_i$). The current sheet is bent by secondary kink instability and form plasmoids. In the case of $L_z=20d_i$ (run B), the elongated current sheet is also unstable to the secondary tearing instability, small secondary magnetic islands appear, expand and merge with the large primary magnetic islands. In all cases this major reconnection site becomes quite extended in the $z$ direction (Fig.~\ref{fig:tearkink3}).

An analysis of the generalized Ohm's law reveals that the reconnection electrical field is mostly from pair pressure tensors at the reconnection sites (Fig.~\ref{force}), consistent with previous results.\citep{bb05,hz07}
The dimensionless reconnection rate $E_R=<E_z/B>$ ($<> $ represents an average over the inflow surface, vertical electrical field component $E_z$ and magnetic field $B$ are measured at the inflow surface) after the establishment of the major reconnection site remains fast and time dependent, with
$E_R \approx 0.068$ and $0.064$ for run B and C, respectively. These rates are smaller than the value in the non-relativistic regime. \cite{ydk08} The reduction of reconnection rate might be due to the higher effective inertia in the relativistic case. \cite{zh07} The inflow surface is chosen to be $\pm 23d_i$ upstream from the dominant reconnection site and has an extent of $12.5d_i$ in $y$. The time-dependent nature of the reconnection rate is due to both the formation and destruction of the plasmoids (similar to the non-relativistic case).

\subsection{Particle Energization: Time and Location}

We now describe the particle energization processes. Given the fact that the 2D run gives a similar amount of total energy conversion as 3D runs
during the tearing stage, we first discuss the magnetic dissipation process associated mostly with the tearing instability in the $\{x, y\}$ plane and assume that the variation along the z direction does not play a major role. During the second stage, for 3D runs with sufficient length in the $z$ direction, the variation along the z direction plays an important role in particle energization (see \S \ref{sec:kink} below).

Fig. \ref{fig:spectrum} shows the particle energy distributions at different times based on two runs, A (panels {\it a, b,} and {\it d}) and C (panel {\it c}). 
Panels {\it a} and {\it c} represent the particle distribution from the whole computational domain of run A and C. Panels {\it b} and {\it d} represent the particle distribution taken from a sampling box with $\Delta x=\Delta y= 2d_i$ at the magnetic island (\emph{b}) and the reconnection site (\emph{d}) of run A. The four curves in each panel represent four characteristic stages of the evolution: the initial state, the end of linear phase, the end of nonlinear interaction stage, and during the secondary kink (for run C), respectively. 
Note that, for panels {\it b} and {\it d}, the total number of particles within the sampling box is changing with time. We have normalized the distributions at different times so that the curves have the same number of particles within each panel.

The initial particle distributions [the black curves in panel (\emph{a, c})] are composed of two parts: one part from the background plasmas and the other from the current sheet plasmas, which has a higher effective temperature because of the drift. On the other hand, the black curves in panel (\emph{b, d}) are mostly composed of the current sheet plasmas only. Over all, a major part of energy conversion takes place between the end of the linear phase and the establishment of the major reconnection site (between \emph{blue} and \emph{green} lines in panels {\it a} and {\it c}), consistent with Fig. \ref{fig:total}. 

Inspecting panel {\it d}, we find that there is some amount of particle acceleration at the reconnection site, as indicated by the {\it blue} curve. This is consistent with the fact that the reconnection electric field is the largest at these locations. But such acceleration, although having large local electric fields, is very limited spatially and does not last in time since the particles can not stay at the reconnection site for a long time but rather quickly leave the reconnection sites via outflows and enter the magnetic islands, as indicated by the {\it green} and {\it red} curves when the particle energy distribution becomes less energetic than the initial distribution. Eventually the plasmas at the reconnection sites are replaced by the edge plasmas from inflow, from the black line with nonzero $\beta$ to the red line with zero $\beta$. (Note that for a given drifting relativistic Maxwellian distribution, the absolute value of the slope is proportional to $\propto(1-\beta)/T$ using the log-linear scale, where $\beta$ is the drift velocity.) 

Panel {\it b}, however, reveals that a large amount of energy conversion occurs in the magnetic island region. Even though the electric field in these island regions might not be as high when compared to the peak reconnection electric field, both the volume and number of particles that experience these electric fields are much larger. The particles are confined inside the magnetic islands for a much longer time than at the reconnection site. Furthermore, by analyzing particle distribution changes within much smaller time spans (see below), we find that the energetic particle population in the island regions must be produced {\it in-situ}, i.e., the majority of the energetic particles do not get energized in the reconnection region then propagate to the island regions. Note that plasmas inside the islands seem to be eventually thermalized into another relativistic Maxwellian distribution with a temperature roughly $2.3$ times the initial temperature of the background plasmas [the green and red lines of Fig.~\ref{fig:spectrum} (\emph{a,b})].

\subsection{Particle Energization: Mechanism}

\begin{figure}[tbp]
\begin{center}
\scalebox{0.8}{\includegraphics{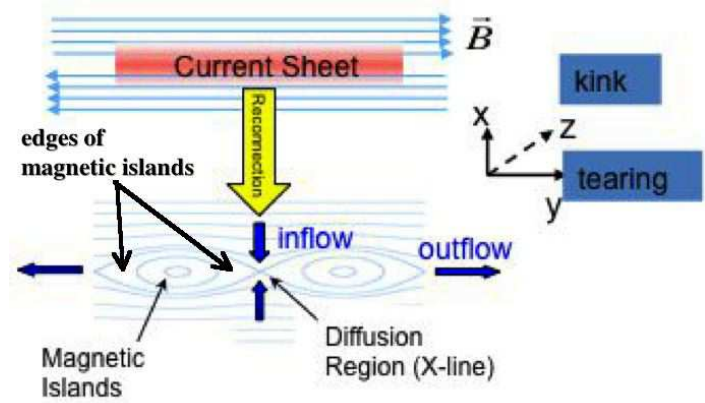}}
\caption{\label{fig:geo} (Color)~Overall geometry of the simulation. Two main types of instabilities are excited in this configuration: the tearing instability along the $k_y$ direction and the kink instability along the $k_z$ direction. 
Initially there is an anti-parallel magnetic field (\emph{cyan arrows}) in the $x$-$y$ plane and a current sheet (\emph{pink}) in the $z$ direction at the interface. After an initial perturbation from thermal noise, magnetic reconnection in $x$-$y$ plane would take place. Inflow (\emph{blue vertical arrows}), outflow (\emph{blue horizontal arrows}) and magnetic islands would form. Different from 2D magnetic reconnection, in the $z$ direction, the strong current induces current-driven instabilities such as kink instability. The edges of magnetic islands are also indicated by \emph{black} arrows.}
\end{center}
\end{figure}

\begin{figure}[tbp]
\begin{center}
\scalebox{0.4}{\includegraphics{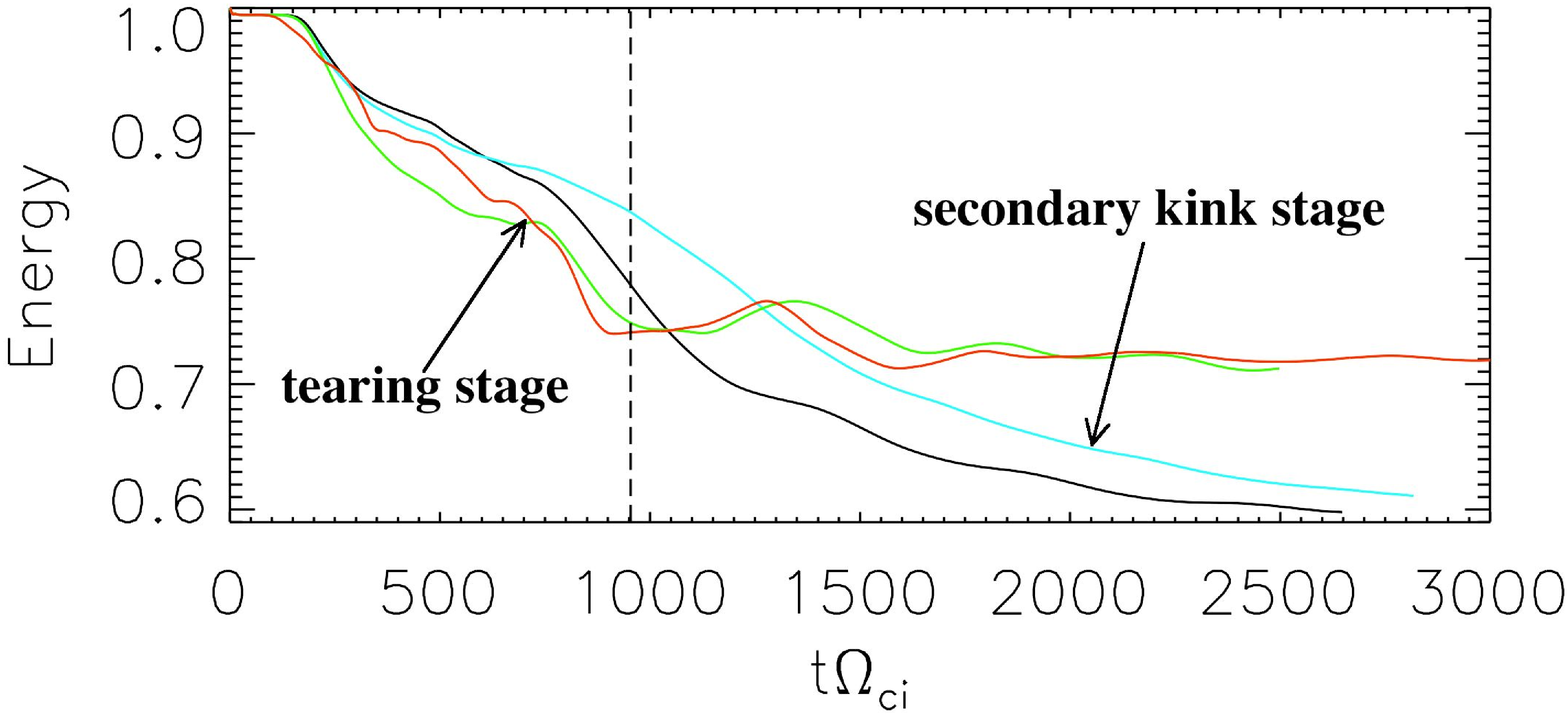}}
\caption{\label{fig:total} (color)~Time evolution of magnetic energy with $L_x=L_y=200d_i$. Red, green, black and blue curves are for 
$L_z=0.195$ (2D), $L_z = 20, 50,$ and $200d_i$, respectively. All energies are normalized to the initial total magnetic energy. The current sheet starts to bend (kink unstable) in the $z$ direction at $t\Omega_c\sim954$ (dash line).}
\end{center}
\end{figure}

\begin{figure}[tbp]
\begin{center}
\scalebox{0.4}{\includegraphics{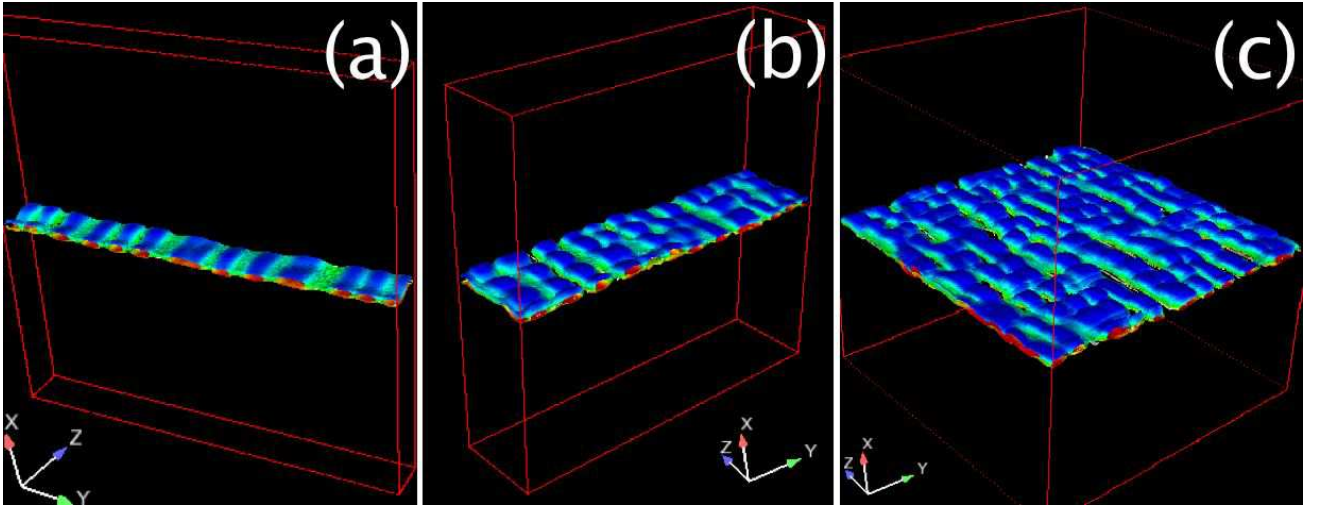}}
\caption{\label{fig:tearkink1} 
(Color) Linear phase of the simulation. Many small magnetic islands resulting from tearing undulate in the $z$ direction due to primary kink instability ($t\Omega_c \lesssim 250$). The blue-red regions are iso-surfaces of density with color indicating peak of $B_y$ (blue: negative; red: positive). These iso-surfaces enclose the volume of the magnetic islands. Embedded are the green regions for the iso-surfaces of weak $|B|$ with color indicating peak of $|J|$. These iso-surfaces enclose the volume of the current sheets at the reconnection sites. Contrary to the non-relativistic limit, the tearing mode appears earlier than the kink modes. (\emph{a}): run B,$L_z=20d_i$; (\emph{b}): run C, $L_z=50d_i$; (\emph{c}): run D, $L_z=200d_i$.}
\end{center}
\end{figure}

\begin{figure}[tbp]
\begin{center}
\scalebox{0.4}{\includegraphics{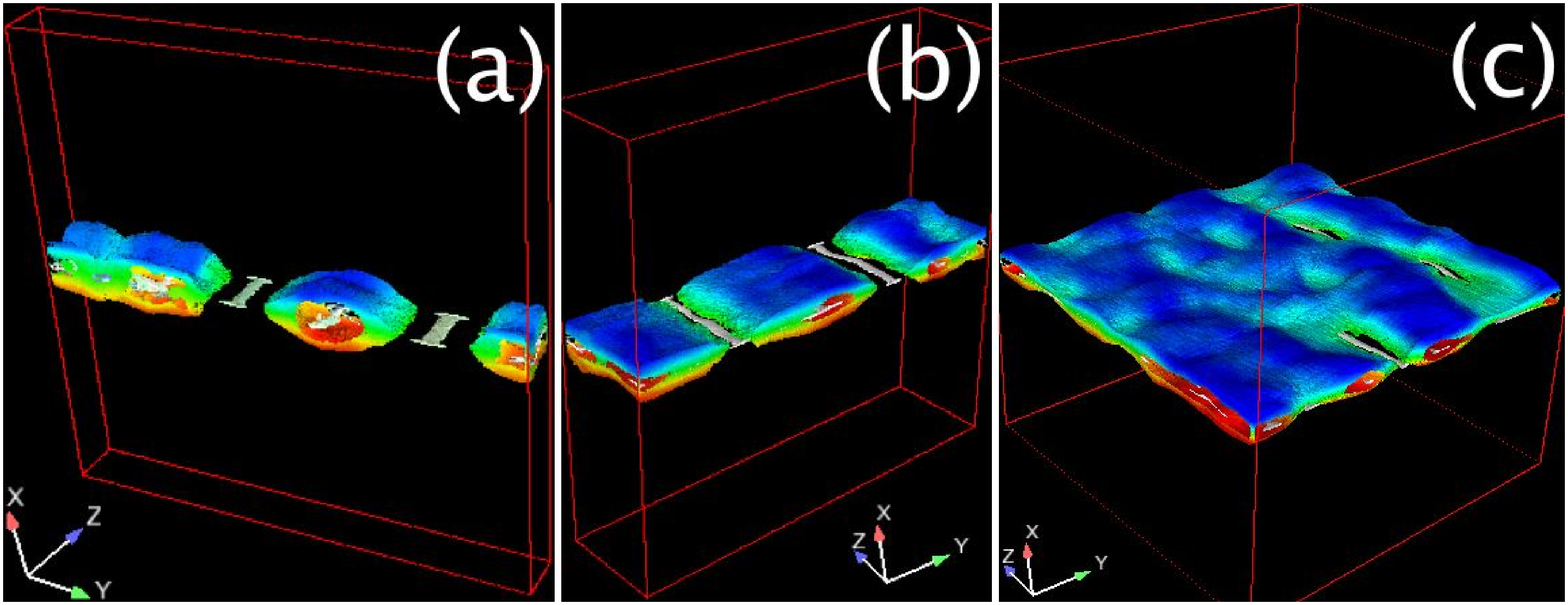}}
\caption{\label{fig:tearkink2} 
(Color) Nonlinear interaction stage of the simulation. The blue-red regions are iso-surfaces of density with color indicating peak of $B_y$ (blue: negative; red: positive). These iso-surfaces enclose the volume of the magnetic islands. Embedded are the white regions for the iso-surfaces of weak $|B|$ with color indicating peak of $|J|$. These iso-surfaces enclose the volume of the current sheets at the reconnection sites. The magnetic islands formed from the tearing instability coalesce to produce a dominant reconnection site through which most of the magnetic flux is processed ($t\Omega_c \lesssim 954$). The localized patchy reconnection sites evolve as they self-organize in $z$ to form a single, large diffusion region. }
\end{center}
\end{figure}

\begin{figure}[!htp]
\begin{center}
\scalebox{0.4}{\includegraphics{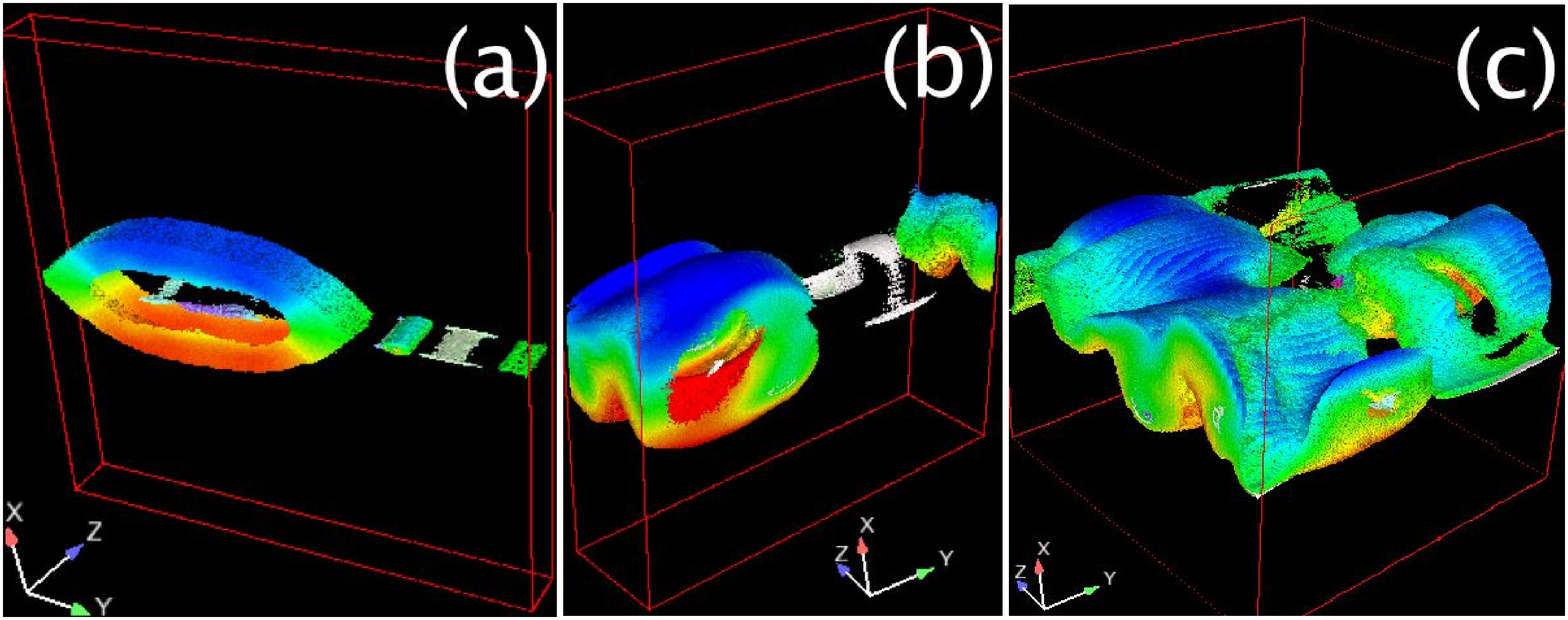}}
\caption{\label{fig:tearkink3} 
(Color) Saturation stage of the simulation. The blue-red regions are iso-surfaces of density with color indicating peak of $B_y$ (blue: negative; red: positive). These iso-surfaces enclose the volume of the magnetic islands. Embedded are the white regions for the iso-surfaces of weak $|B|$ with color indicating peak of $|J|$. These iso-surfaces enclose the volume of the current sheets at the reconnection sites. The dominant diffusion region expands in $y$ to form a highly elongated current sheet unstable to the secondary kink instability. The current sheet is bent by the secondary kink instability and form plasmoids ($t\Omega_c \gtrsim 954$).}
\end{center}
\end{figure}

\begin{figure}[!htp]
\begin{center}
 \scalebox{0.4}{\includegraphics{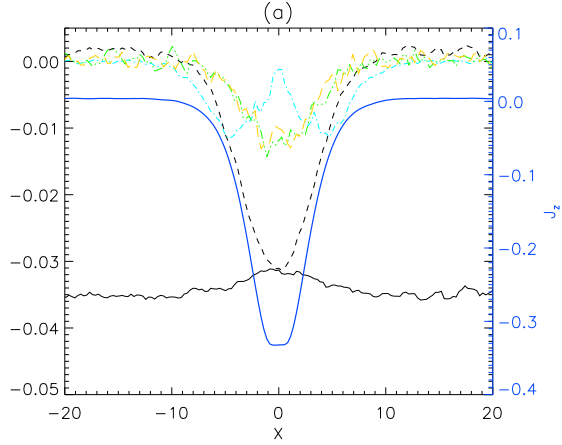}}
  \caption{\label{force}(color)~$x$ distribution of $\mathbf{E}_z$ (black solid), $[\mathbf{E}+\mathbf{V}\times\mathbf{B}/c]_z$ (black dash), electron pressure term $[-\frac{1}{2ne}\grad\cdot \mathbf{P}_e]_z$ (green), positron pressure term $[\frac{1}{2ne}\grad\cdot \mathbf{P}_i]_z$ (yellow), convective bulk inertial term $[\frac{m}{2ne^2}\grad\cdot(\mathbf{J}\mathbf{v}+\mathbf{v}\mathbf{J})]_z$ (cyan) and $J_z$ (blue) at an X point $(x,y,z)=(0,232,0)$ of run B at $t\Omega_{\rm ci}\sim1011$. The time dependent term $[\frac{m}{2ne^2}\frac{\partial \mathbf{J}}{\partial t}]_z$, which is hard to calculate accurately, is not shown here.\cite{bb05} All terms are normalized by $B_0V_{\rm A}/c$, where $V_{\rm A}=B_0/[4\pi n_0(m_e+m_i)]$. $J_z$ is normalized by $n_0eV_{\rm A}$. $x$ is normalized by the half thickness of the initial current sheet $L$.
}
\end{center}
\end{figure}

Although particle acceleration is expected at the reconnection sites, it is not so obvious why the majority of particle energization occurs in the island regions in these simulations. During the tearing stage, outflows from reconnection (cf. Fig. \ref{fig:geo}) lead to strong and $y$-dependent $\vec{E}=-(\vec{V}\times\vec{B})/c$ at the edges of islands (see Fig.~\ref{fig:geo}), where $\vec{V}$ is mostly along the y-direction and $\vec{B}$ is in the $\{x, y\}$ plane. The resulting electric field component $E_z$ is quite significant because the outflow velocity can become quasi-relativistic. This is a peculiarity of the relativistic pair plasmas because of the absence of ions, which are much heavier than the electrons. Therefore the Alfven velocity (the typical outflow speed for particles) can become relativistic too, leading to a large component of $-V\times B$.\cite{jtls04} We find that these electric fields tend to be long-lived and are distributed over large spatial volumes. 
Consequently they are responsible for the majority of magnetic energy conversion to particles, at least for the configurations we have studied here. 

To understand the details of particle energization, we plot the spatial distribution of the normalized $E^2-B^2$ in Fig. \ref{fig:acceleration}, using results from run A. The reconnection regions indeed have $E^2-B^2 > 0$, indicating the existence of a net electric field that can accelerate particles. Most other regions have $E^2 - B^2 < 0$, which has been usually interpreted as not useful for particle acceleration because there exists a Lorentz transformation that makes electric field vanish. \citep{hz07} Such consideration, however, does not apply when there are large spatial variations of $E_z$ or $B_x$ or both. In the bottom panel of Fig. \ref{fig:acceleration}, we plot the $B_x$ and $E_z$ profiles along the $y$-axis (with $x=0$). As particles try to gyrate in the $\{y,z\}$ plane due to $B_x$ but experience a spatially varying $E_z$ along the $y$-direction within their gyro-orbits, net acceleration can occur.

Particular types of field structures are favored in particle acceleration as shown in Fig.~\ref{fig:acceleration2}. Panel (a) emphasizes the spatial variations in $E_z$. Particles gain more energy in the larger $|E_z|$ side than the energy lost in the smaller $|E_z|$ side. Another scenario emphasizes the variations in $B_x$. Particles have larger gyroradius in the smaller $|B_x|$ side than in the larger $|B_x|$ side, leading to net particle acceleration in the $z$ direction.\cite{hmts01} Fig. \ref{fig:acceleration} (\emph{bottom}) shows that both of these structures are present when the tearing instability becomes nonlinear. To confirm this acceleration process,  we have performed test particle calculations using the initial particle distribution as in the simulation \cite{ks03,zh07} and with three different field structures: 1) constant $B_x$ but spatially varying $E_z$ [Fig.~\ref{fig:acceleration2} (\emph{a})]; 2) constant $E_z$ but spatially varying $B_x$; 3) temporally and spatially varying $B_x$ and $E_z$ directly from the 2D simulation run A [Fig.\ref{fig:acceleration2} (\emph{b})]. The $E_z$ and $B_y$ profile data from the nonlinear simulation are fitted with first order polynomials by the least square fit method at $t\Omega_c\sim329$ between $y=60d_i$ and $y=110d_i$. And then a spatially constant $E_z=0.1$ and a realistic $B_y$ profile are used for (1) and a spatially constant $B_y=1.0$ and a realistic $E_z$ profile are used for (2).The total number of particles used in these test particle calculations is around $500,000$. We find that the varied $E_z$ scheme [Fig.~\ref{fig:acceleration2} (a)] produces $\sim10$ times more efficient particle acceleration than the varied $B_x$ scheme. Quantitatively, for efficient acceleration, it is necessary for the typical $E_z$ variation scale to be smaller than the particle's gyroradius: 
\begin{equation}
\label{eq1}
|\widetilde{E}_z/(\partial E_z/\partial y)|\lesssim\gamma mc^2/(e|B_x|)\,,
\end{equation} 
where $\widetilde{E}_z$ is the characteristic $E_z$ along the particle trajectories. 
This suggests that the more sharply varied $E_z$ or weaker $B_x$ leads to more efficient particle acceleration, which is confirmed by our test particle calculations [Fig.~\ref{fig:acceleration2} (\emph{c})]. Furthermore, test particle calculation using the time-dependent field output from the simulation run A gives $d\gamma/d(t\Omega_c)\sim0.02$ by statistical average [Fig.~\ref{fig:acceleration2} (\emph{d})], which is sufficient to explain the particle energy gain before the secondary kink instability kicks in.  We confirm that 
particles are mostly accelerated at the edges of the islands (e.g., the region around $y \sim 80$ and $90 d_i$, also see Fig.~\ref{fig:geo}). This type of field configuration is quite robust and is found to persist throughout the whole tearing stage.  
The long-term ($t\Omega_c\gtrsim500$) test particle calculations result in quasi-Maxwellian distributions at low energy, but with significant $(\gtrsim15\%)$ non-thermal high-energy particles ($\gamma \gtrsim20$) at $t\Omega_c\gtrsim750$ [Fig.~\ref{fig:acceleration2} (\emph{d})], which is not observed in the real simulations. This is most likely due to the thermalization process in real simulations (see below).

\subsection{Particle Energization: Thermalization}\label{thermalization}

The test particle study shown in Fig. \ref{fig:acceleration2} suggests that the fields from the tearing instability could accelerate particles to high energies ($\gamma \sim 100$). Fig. \ref{fig:thermalization}, however, indicates that the non-thermal high-energy particles from this acceleration process have been significantly thermalized. The particle distributions are taken from a sampling box at $x = 0$ and $y=89 d_i$, which is near the right edge of the central island shown in Fig. \ref{fig:acceleration}. Specifically, as electrons are accelerated in the $+ p_z$ direction (panel {\it b}
of Fig. \ref{fig:thermalization}), their $+ p_z$ momentum turns into $-p_y$ momentum due to the relatively large $B_x$, e.g., the extended tails in
$- p_y$ shown as unsymmetric black dash line in panel {\it a}. Note that the positrons will experience similar processes, except that they have $- p_z$ turning into $- p_y$.  Subsequent evolution shows that there exists a series of plateaus with decaying magnitudes at the $- p_y$ side of particle number distribution [blue and green lines of Fig.~\ref{fig:thermalization} (\emph{a})]. The particle distribution then turns into an approximately symmetric Maxiwellian distribution in $p_y$ as shown by the red line of Fig.~\ref{fig:thermalization} (\emph{a}). This distribution can be fitted with two temperatures, suggesting that the energy gained through the $E_z$ acceleration has been thermalized in the $p_y$ direction.

Note that the time difference between the black dash and red curves in Fig. \ref{fig:thermalization} ({\it a}) is $\delta t\Omega_c \sim 50$, this implies that the thermalization process is quite efficient. We suggest that a mixture of bump-on-tail and two-stream instabilities can cause wave excitation and Landau damping, and this eventually results in a symmetric distribution in $p_y$.  
This process could be fast because the strength of two-stream instability and Landau-damping is proportional to plasma frequency $\omega_{p}$ in pair plasmas , much larger than in electron-ion plasmas ($\sim\omega_{pe}\sqrt{m_e/m_i}$). The above particle thermalization process would happen repeatedly inside the islands and eventually all the plasmas inside the islands are thermalized into a higher temperature. From Fig.~\ref{fig:thermalization} (\emph{a}), the typical thermalization time scale is found to be $\delta t\Omega_c\sim50$.

It is worth noting that during the time window shown in Fig. \ref{fig:thermalization}, no island merging is observed to take place. 
Furthermore, similar particle distribution evolution shown in Fig.~\ref{fig:thermalization} appears also at both downstream ($y=87d_i$) and upstream $(y=91d_i)$ regions simultaneously around $y=89d_i$ region. This rules out the possibility that this energization is due to the particle convection, but supports the suggestion that the thermalization is due to local ``micro-instabilities". In addition, we observe that particles are drifting slowly in the $y$ direction, with a typical \emph{particle} velocity $v_y/c = p_y/(\gamma m) \lesssim 0.2$ (not \emph{phase} velocity). This means that the particles travel $\lesssim 5 d_i$ in $\delta t \Omega_c \sim 50$ in the $y$-direction, which is much smaller than the size of the magnetic islands ($\sim 30 d_i$), thus particles do not have enough time to bounce between the edges of the islands. The typical gyro-radii of the particles are always $5-7$ times smaller than the island size at every stage, thus the particle acceleration from bouncing between ``walls" of contracting magnetic islands \cite{dscs06} due to the fast gyro-motion is not possible either.

We also find that particles have very small $p_x$ throughout the evolution and the $p_z$ distribution remains ``skewed'' due to the presence of $E_z$ until quite late time ($t\Omega_c \sim 1800$) when the distribution in $p_z$ becomes symmetric as well [Fig. \ref{fig:thermalization} ({\it b})]. 

Even though we mostly present the results from run A, we found similar results with run C. Both the acceleration mechanism that relies on the spatial variation of $E_z$ and the subsequent thermalization process seem to be robust processes for particle energization in pair plasmas.

\subsection{Particle Energization: The Role of the Secondary Kink Instability}
\label{sec:kink}

For both run C and D, there is additional particle energization in the secondary kink stage. 
We find that current sheets along the $z$ direction emanating from both the reconnection and island regions undergo the secondary kink instability with similar wavelength $\sim40d_i$ (Fig.~\ref{fig:second_kink}). The magnitude of kinking at the island region is larger because the kinking at the reconnection site is suppressed by the reconnection inflow (Fig.~\ref{fig:second_kink}). The onset of the secondary kink instability inside the island is also delayed compared to the one at the reconnection site since the the current sheet at the magnetic islands is thicker. Those secondary kink instabilities are new kink modes since the the initial primary kink modes have been stabilized during the tearing stage.

The appearance of the $z$-dependent $B_y$ and $E_z$ resulting from the secondary kink instability (Fig.~\ref{fig:second_kink}) looks very similar to Fig.~1 of \citet{zh05}, which suggests that the same particle acceleration mechanism works for the secondary kink instability during this stage, \emph{i.e.}, a ``DC" acceleration channel mechanism (for detailed discussion please see \citet{zh05}). 
This same mechanism accelerates high-energy particles ($\gamma\gtrsim10$) efficiently at both magnetic islands and reconnection sites. This leads to the production of the non-thermal particles [high energy tail of \emph{red} line of Fig.~\ref{fig:spectrum} (\emph{c})], although those non-thermal particles would be mostly thermalized by the above thermalization process associated with $B_x$ and only some of non-thermal particles might be left. Here $E_z\sim0.2$ is mostly due to the charge separation from the relative electron and positron drift, \cite{zh07} which leads to the onset of the secondary kink instability. 

The high-energy particles undergo a fast streaming motion in the $z$ direction but at the same time executing frequent cyclotron motion in the $(p_y,p_z)$ space, leading to efficient thermalization as discussed in Sec.~\ref{thermalization}. Such a strong thermalization process is not observed in previous 2D kink studies \citep{zh05} since $B_x\sim1$ induced from tearing modes in this study is much larger than in those cases $\sim0.1$. This might explain a relatively higher percentage of non-thermal particles found in those studies \cite{zh05}. 

It is worth emphasizing that, at least for the
initial configuration we have studied here, the particle
energization by tearing occurs mostly inside the magnetic
islands and produces mostly thermal heating in both stages.
But particle energization by the secondary kink instability
leads to some non-thermal particles in the second kink stage.
These effects could only be studied with 3D simulation of
a large enough computational domain in every direction.

\section{Summary}\label{conclusion}

In this paper we have presented large scale 3D particle-in-cell (PIC) simulations to examine particle energization in magnetic reconnection of relativistic electron-positron (pair) plasmas. These simulations are large enough to accommodate a sufficient number of tearing and kink modes. The interplay of these two main instabilities regulates the overall evolution: from the initial linear tearing and kink growth with the appearance of small magnetic islands, to 
the merging of small magnetic islands into a big one, establishing the major reconnection sites, to the elongated current sheet becoming 
unstable to the secondary kink instability, forming plasmoids. The particle energization could be divided into two main stages: (1) the tearing stage ($t\Omega_c\lesssim954)$, and (2) the secondary kink stage ($t\Omega_c\gtrsim954)$. We find that particles are mostly energized inside the magnetic islands during the tearing stage due to the spatially varying electric fields produced by the outflows from reconnection. Secondary kink instability also leads to some particle acceleration. Accelerated particles are,
however,  observed to be thermalized. Since
the physical conditions for both 
the energization process and the thermalization process discussed in this study are quite typical in relativistic pair reconnection, we suggest that these processes are germane in understanding the magnetic energy dissipation in pair plasmas. 

In a relativistic current sheet, previous literature reported that the linear kink instability is faster than the linear tearing instability unless the reconnection is driven.\cite{zh05,zh05b,zh07,zh08} In this work, contrary to the above well-known results, the linear tearing instability is faster than the linear kink instability, at least in our specific parameters. This result raises another hint to the striped wind problem.\cite{ks03}  It is worth of emphasizing that all the calculations reported in this paper are done with an initial anti-parallel magnetic field without a guide field for pair plasmas. The introduction of a uniform guide field would significantly change the results as reported in \citet{hz07}. Compared to a relativistic reconnection without a guide field, the reconnection rate is slower with a guide field due to a lower compressibility effect of the guide magnetic field. More importantly with a guide field, the particle acceleration is mostly done at the reconnection site, whereas it is inside the magnetic island without a guide field.\cite{hz07} The magnetic field configuration in the real astrophysical situation such as pulsar wind would be even more complicated. We need to be cautious when trying to apply the particle energization processes reported in this paper to the real astrophysical situations. Additionally the energization process in pair plasmas may find only limited applications
in the  solar or magnetospheric environments  where electron-ion reconnection should be dominant.

\begin{acknowledgments}
This work was supported by the U.S. Department of Energy and was funded in part by the DoE Office of Fusion Energy Science, a grant from IGPP and the
LDRD program at the Los Alamos National Laboratory with contract no. DE-AC52-06NA25396.
\end{acknowledgments}


\begin{thebibliography}{20}
\expandafter\ifx\csname natexlab\endcsname\relax\def\natexlab#1{#1}\fi
\expandafter\ifx\csname bibnamefont\endcsname\relax
  \def\bibnamefont#1{#1}\fi
\expandafter\ifx\csname bibfnamefont\endcsname\relax
  \def\bibfnamefont#1{#1}\fi
\expandafter\ifx\csname citenamefont\endcsname\relax
  \def\citenamefont#1{#1}\fi
\expandafter\ifx\csname url\endcsname\relax
  \def\url#1{\texttt{#1}}\fi
\expandafter\ifx\csname urlprefix\endcsname\relax\def\urlprefix{URL }\fi
\providecommand{\bibinfo}[2]{#2}
\providecommand{\eprint}[2][]{\url{#2}}

\bibitem[{\citenamefont{Zenitani and Hoshino}(2005{\natexlab{a}})}]{zh05}
\bibinfo{author}{\bibfnamefont{S.}~\bibnamefont{Zenitani}} \bibnamefont{and}
  \bibinfo{author}{\bibfnamefont{M.}~\bibnamefont{Hoshino}},
  \bibinfo{journal}{ApJ} \textbf{\bibinfo{volume}{618}}, \bibinfo{pages}{L111}
  (\bibinfo{year}{2005}{\natexlab{a}}).

\bibitem[{\citenamefont{Kirk and Skj{\ae}raasen}(2003)}]{ks03}
\bibinfo{author}{\bibfnamefont{J.~G.} \bibnamefont{Kirk}} \bibnamefont{and}
  \bibinfo{author}{\bibnamefont{Skj{\ae}raasen}}, \bibinfo{journal}{ApJ}
  \textbf{\bibinfo{volume}{591}}, \bibinfo{pages}{366} (\bibinfo{year}{2003}).

\bibitem[{\citenamefont{Bessho and Bhattacharjee}(2005)}]{bb05}
\bibinfo{author}{\bibfnamefont{N.}~\bibnamefont{Bessho}} \bibnamefont{and}
  \bibinfo{author}{\bibfnamefont{A.}~\bibnamefont{Bhattacharjee}},
  \bibinfo{journal}{PRL} \textbf{\bibinfo{volume}{95}},
  \bibinfo{pages}{245001} (\bibinfo{year}{2005}).

\bibitem[{\citenamefont{Chen et~al.}(2009)\citenamefont{Chen, Wilks, Bonlie,
  Liang, Myatt, Price, Meyerhofer, and Beiersdorfer}}]{cwb09}
\bibinfo{author}{\bibfnamefont{H.}~\bibnamefont{Chen}},
  \bibinfo{author}{\bibfnamefont{S.~C.} \bibnamefont{Wilks}},
  \bibinfo{author}{\bibfnamefont{J.~D.} \bibnamefont{Bonlie}},
  \bibinfo{author}{\bibfnamefont{E.~P.} \bibnamefont{Liang}},
  \bibinfo{author}{\bibfnamefont{J.}~\bibnamefont{Myatt}},
  \bibinfo{author}{\bibfnamefont{D.~F.} \bibnamefont{Price}},
  \bibinfo{author}{\bibfnamefont{D.~D.} \bibnamefont{Meyerhofer}},
  \bibnamefont{and}
  \bibinfo{author}{\bibfnamefont{P.}~\bibnamefont{Beiersdorfer}},
  \bibinfo{journal}{PRL} \textbf{\bibinfo{volume}{102}},
  \bibinfo{eid}{105001} (\bibinfo{year}{2009}).


\bibitem[{\citenamefont{Zenitani and Hoshino}(2001)}]{zh01}
\bibinfo{author}{\bibfnamefont{S.}~\bibnamefont{Zenitani}} \bibnamefont{and}
  \bibinfo{author}{\bibfnamefont{M.}~\bibnamefont{Hoshino}},
  \bibinfo{journal}{ApJ} \textbf{\bibinfo{volume}{562}}, \bibinfo{pages}{L63}
  (\bibinfo{year}{2001}).

\bibitem[{\citenamefont{Bessho and Bhattacharjee}(2007)}]{bb07}
\bibinfo{author}{\bibfnamefont{N.}~\bibnamefont{Bessho}} \bibnamefont{and}
  \bibinfo{author}{\bibfnamefont{A.}~\bibnamefont{Bhattacharjee}},
  \bibinfo{journal}{PoP} \textbf{\bibinfo{volume}{14}}, \bibinfo{pages}{056503}
  (\bibinfo{year}{2007}).

\bibitem[{\citenamefont{Zenitani and Hoshino}(2007)}]{zh07}
\bibinfo{author}{\bibfnamefont{S.}~\bibnamefont{Zenitani}} \bibnamefont{and}
  \bibinfo{author}{\bibfnamefont{M.}~\bibnamefont{Hoshino}},
  \bibinfo{journal}{ApJ} \textbf{\bibinfo{volume}{670}}, \bibinfo{pages}{702}
  (\bibinfo{year}{2007}).

  \bibitem[{\citenamefont{Jaroschek et~al.}(2004)\citenamefont{Jaroschek,
  Treumann, Lesch, and Scholer}}]{jtls04}
\bibinfo{author}{\bibfnamefont{C.~H.} \bibnamefont{Jaroschek}},
  \bibinfo{author}{\bibfnamefont{R.~A.} \bibnamefont{Treumann}},
  \bibinfo{author}{\bibfnamefont{H.}~\bibnamefont{Lesch}}, \bibnamefont{and}
  \bibinfo{author}{\bibfnamefont{M.}~\bibnamefont{Scholer}},
  \bibinfo{journal}{PoP} \textbf{\bibinfo{volume}{11}},
  \bibinfo{pages}{1151} (\bibinfo{year}{2004}).


\bibitem[{\citenamefont{Zenitani and Hoshino}(2005{\natexlab{b}})}]{zh05b}
\bibinfo{author}{\bibfnamefont{S.}~\bibnamefont{Zenitani}} \bibnamefont{and}
  \bibinfo{author}{\bibfnamefont{M.}~\bibnamefont{Hoshino}},
  \bibinfo{journal}{PRL} \textbf{\bibinfo{volume}{95}}, \bibinfo{pages}{095001}
  (\bibinfo{year}{2005}{\natexlab{b}}).
  
  
\bibitem[{\citenamefont{Zenitani and Hoshino}(2005{\natexlab{b}})}]{zh08}
\bibinfo{author}{\bibfnamefont{S.}~\bibnamefont{Zenitani}} \bibnamefont{and}
  \bibinfo{author}{\bibfnamefont{M.}~\bibnamefont{Hoshino}},
  \bibinfo{journal}{ApJ} \textbf{\bibinfo{volume}{677}}, \bibinfo{pages}{530}
  (\bibinfo{year}{2005}{\natexlab{b}}).

\bibitem[{\citenamefont{Yoon et~al.}(1996)\citenamefont{Yoon, Drake, and
  Anthony}}]{yda96}
\bibinfo{author}{\bibfnamefont{P.~H.} \bibnamefont{Yoon}},
  \bibinfo{author}{\bibfnamefont{J.~F.} \bibnamefont{Drake}}, \bibnamefont{and}
  \bibinfo{author}{\bibfnamefont{T.~Y.} \bibnamefont{Anthony}},
  \bibinfo{journal}{JGR} \textbf{\bibinfo{volume}{101}},
  \bibinfo{pages}{27327} (\bibinfo{year}{1996}).

\bibitem[{\citenamefont{Shinohara et~al.}(2001)\citenamefont{Shinohara, Suzuki,
  Fujimoto, and Hoshino}}]{ssf01}
\bibinfo{author}{\bibfnamefont{I.}~\bibnamefont{Shinohara}},
  \bibinfo{author}{\bibfnamefont{H.}~\bibnamefont{Suzuki}},
  \bibinfo{author}{\bibfnamefont{M.}~\bibnamefont{Fujimoto}}, \bibnamefont{and}
  \bibinfo{author}{\bibfnamefont{M.}~\bibnamefont{Hoshino}},
  \bibinfo{journal}{PRL} \textbf{\bibinfo{volume}{8709}},
  \bibinfo{pages}{5001} (\bibinfo{year}{2001}).

\bibitem[{\citenamefont{Daughton}(1998)}]{dw98}
\bibinfo{author}{\bibfnamefont{W.}~\bibnamefont{Daughton}},
  \bibinfo{journal}{JGR} \textbf{\bibinfo{volume}{103}},
  \bibinfo{pages}{29429} (\bibinfo{year}{1998}).

\bibitem[{\citenamefont{Hesse and Zenitani}(2007)}]{hz07}
\bibinfo{author}{\bibfnamefont{M.}~\bibnamefont{Hesse}} \bibnamefont{and}
  \bibinfo{author}{\bibfnamefont{S.}~\bibnamefont{Zenitani}}, \bibinfo{journal}{PoP}
  \textbf{\bibinfo{volume}{14}}, \bibinfo{pages}{112102}
  (\bibinfo{year}{2007}).
  
  

\bibitem[{\citenamefont{Yin et~al.}(2008)\citenamefont{Yin, Daughton,
  Karimabadi, Albright, Bowers, and Margulies}}]{ydk08}
\bibinfo{author}{\bibfnamefont{L.}~\bibnamefont{Yin}},
  \bibinfo{author}{\bibfnamefont{W.}~\bibnamefont{Daughton}},
  \bibinfo{author}{\bibfnamefont{H.}~\bibnamefont{Karimabadi}},
  \bibinfo{author}{\bibfnamefont{B.~J.} \bibnamefont{Albright}},
  \bibinfo{author}{\bibfnamefont{K.~J.} \bibnamefont{Bowers}},
  \bibnamefont{and}
  \bibinfo{author}{\bibfnamefont{J.}~\bibnamefont{Margulies}},
  \bibinfo{journal}{PRL} \textbf{\bibinfo{volume}{101}},
  \bibinfo{pages}{125001} (\bibinfo{year}{2008}).

\bibitem[{\citenamefont{Hoshino et~al.}(2001)\citenamefont{Hoshino, Mukai,
  Teresawa, and Shinohara}}]{hmts01}
\bibinfo{author}{\bibfnamefont{M.}~\bibnamefont{Hoshino}},
  \bibinfo{author}{\bibfnamefont{T.}~\bibnamefont{Mukai}},
  \bibinfo{author}{\bibfnamefont{T.}~\bibnamefont{Teresawa}}, \bibnamefont{and}
  \bibinfo{author}{\bibfnamefont{I.}~\bibnamefont{Shinohara}},
  \bibinfo{journal}{JGR} \textbf{\bibinfo{volume}{106}},
  \bibinfo{pages}{25979} (\bibinfo{year}{2001}).

\bibitem[{\citenamefont{Pritchett}(2006)}]{pp06}
\bibinfo{author}{\bibfnamefont{P.~L.} \bibnamefont{Pritchett}},
  \bibinfo{journal}{GRL} \textbf{\bibinfo{volume}{33}},
  \bibinfo{pages}{L13104} (\bibinfo{year}{2006}).

\bibitem[{\citenamefont{Chen et~al.}(2008/01//print)\citenamefont{Chen,
  Bhattacharjee, Puhl-Quinn, Yang, Bessho, Imada, Muhlbachler, Daly, Lefebvre,
  Khotyaintsev et~al.}}]{cbp08}
\bibinfo{author}{\bibfnamefont{L.~J.} \bibnamefont{Chen}},
  \bibinfo{author}{\bibfnamefont{A.}~\bibnamefont{Bhattacharjee}},
  \bibinfo{author}{\bibfnamefont{P.~A.} \bibnamefont{Puhl-Quinn}},
  \bibinfo{author}{\bibfnamefont{H.}~\bibnamefont{Yang}},
  \bibinfo{author}{\bibfnamefont{N.}~\bibnamefont{Bessho}},
  \bibinfo{author}{\bibfnamefont{S.}~\bibnamefont{Imada}},
  \bibinfo{author}{\bibfnamefont{S.}~\bibnamefont{Muhlbachler}},
  \bibinfo{author}{\bibfnamefont{P.~W.} \bibnamefont{Daly}},
  \bibinfo{author}{\bibfnamefont{B.}~\bibnamefont{Lefebvre}},
  \bibinfo{author}{\bibfnamefont{Y.}~\bibnamefont{Khotyaintsev}},
  \bibinfo{author}{\bibfnamefont{A.}~\bibnamefont{Fazakerley}}, \bibnamefont{and}
  \bibinfo{author}{\bibfnamefont{E.}~\bibnamefont{Georgescu}},
  \bibinfo{journal}{Nat. Phys.}
  \textbf{\bibinfo{volume}{4}}, \bibinfo{pages}{19}
  (\bibinfo{year}{2008}),
  
    \bibitem[{\citenamefont{Dmitruk, Matthaeus and Seenu}(2004)}]{dms04}
\bibinfo{author}{\bibfnamefont{P.}~\bibnamefont{Dmitruk}}, 
  \bibinfo{author}{\bibfnamefont{W. H.}~\bibnamefont{Matthaeus}} \bibnamefont{and} \bibinfo{author}{\bibfnamefont{N.}~\bibnamefont{Seenu}}, \bibinfo{journal}{ApJ}
  \textbf{\bibinfo{volume}{617}}, \bibinfo{pages}{667}
  (\bibinfo{year}{2004}).


\bibitem[{\citenamefont{Drake et~al.}(2006)\citenamefont{Drake, Swisdak, Che,
  and SHay}}]{dscs06}
\bibinfo{author}{\bibfnamefont{J.~F.} \bibnamefont{Drake}},
  \bibinfo{author}{\bibfnamefont{M.}~\bibnamefont{Swisdak}},
  \bibinfo{author}{\bibfnamefont{H.}~\bibnamefont{Che}}, \bibnamefont{and}
  \bibinfo{author}{\bibfnamefont{M.~A.} \bibnamefont{Shay}},
  \bibinfo{journal}{Nature} \textbf{\bibinfo{volume}{443}},
  \bibinfo{pages}{553} (\bibinfo{year}{2006}).
  
   
  \bibitem[{\citenamefont{P\'{e}tri and Kirk}(2007{\natexlab{a}})}]{pk07}
\bibinfo{author}{\bibfnamefont{J.}~\bibnamefont{P\'{e}tri}} \bibnamefont{and}
  \bibinfo{author}{\bibfnamefont{J.~G.} \bibnamefont{Kirk}},
  \bibinfo{journal}{Plasma Phys. Control. Fusion}
  \textbf{\bibinfo{volume}{49}}, \bibinfo{pages}{1885}
  (\bibinfo{year}{2007}{\natexlab{a}}).

\bibitem[{\citenamefont{P\'{e}tri and Kirk}(2007{\natexlab{b}})}]{pk07b}
\bibinfo{author}{\bibfnamefont{J.}~\bibnamefont{P\'{e}tri}} \bibnamefont{and}
  \bibinfo{author}{\bibfnamefont{J.~G.} \bibnamefont{Kirk}},
  \bibinfo{journal}{Plasma Phys. Control. Fusion}
  \textbf{\bibinfo{volume}{49}}, \bibinfo{pages}{297}
  (\bibinfo{year}{2007}{\natexlab{b}}).
  
  \bibitem[{\citenamefont{Daughton}(1999)}]{dw99}
\bibinfo{author}{\bibfnamefont{W.}~\bibnamefont{Daughton}},
  \bibinfo{journal}{Phys. Plasmas} \textbf{\bibinfo{volume}{6}},
  \bibinfo{pages}{1329} (\bibinfo{year}{1999}).
  


\end{thebibliography}

\newpage

\begin{figure}[htbp]
\begin{center}
\scalebox{0.4}{\includegraphics{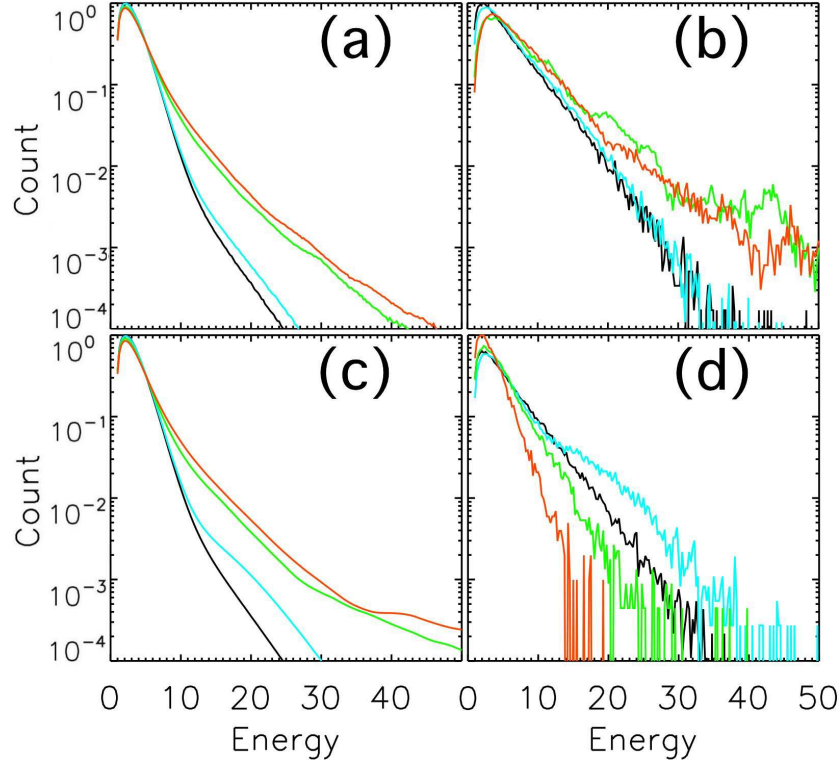}}
\caption{\label{fig:spectrum} (color)~Particle number distribution \emph{v.s.} energy at some typical stages of run A and C. For run A, panel ({\it a}), ({\it b}) 
and ({\it d}) represent the particle distributions from the whole computation domain, at an magnetic island, and at a reconnection site, respectively. For run C, panel ({\it c}) represents the particle distribution from the whole computation domain. All curves are normalized to have the same total number of particles. For (\emph{a,b,d}),  curves \emph{black},~$t\Omega_c=0$;  \emph{blue},~$t\Omega_c=149$; \emph{green},~$t\Omega_c=808$; \emph{red},~$t\Omega_c=1811$. For (\emph{c}), curves \emph{black},~$t\Omega_c=0$;  \emph{blue},~$t\Omega_c=238$; \emph{green},~$t\Omega_c=954$; \emph{red},~$t\Omega_c=1470$.
}
\end{center}
\end{figure}

\begin{figure}[tbp]
\begin{center}
\scalebox{0.4}{\includegraphics{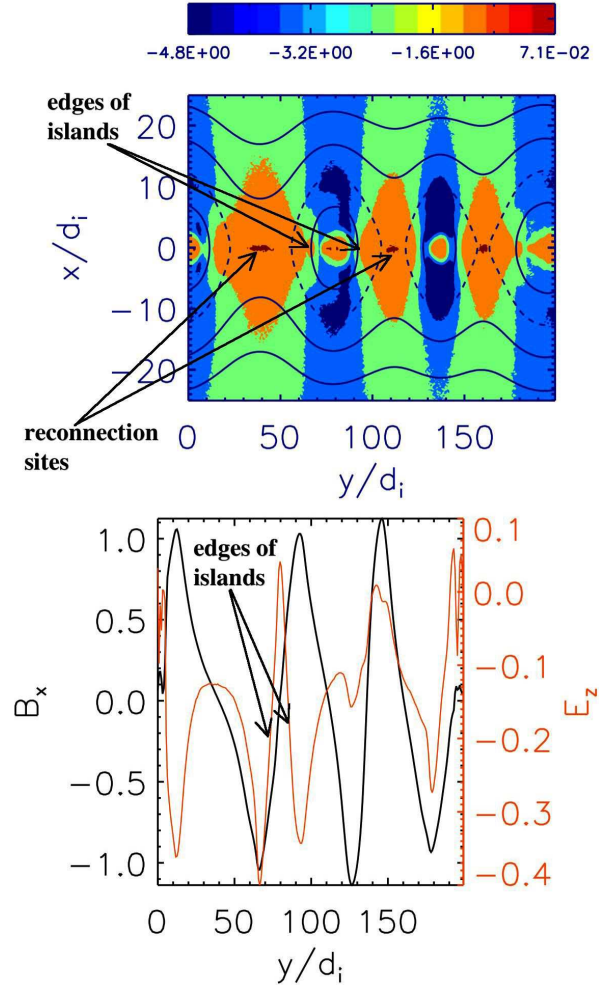}}
\caption{\label{fig:acceleration}~(Color) Contour plot of magnetic flux function $\Psi=-\int B_ydx$ colored by $E^2-B^2$. The \emph{dash} contour lines of magnetic flux function $\Psi$ defines the magnetic islands structures. Three magnetic islands are observed at this moment. Fig.~\ref{fig:acceleration2} focuses on the middle island.
  }
\end{center}
\end{figure}

\begin{figure}[tbp]
\begin{center}
\scalebox{0.3}{\includegraphics{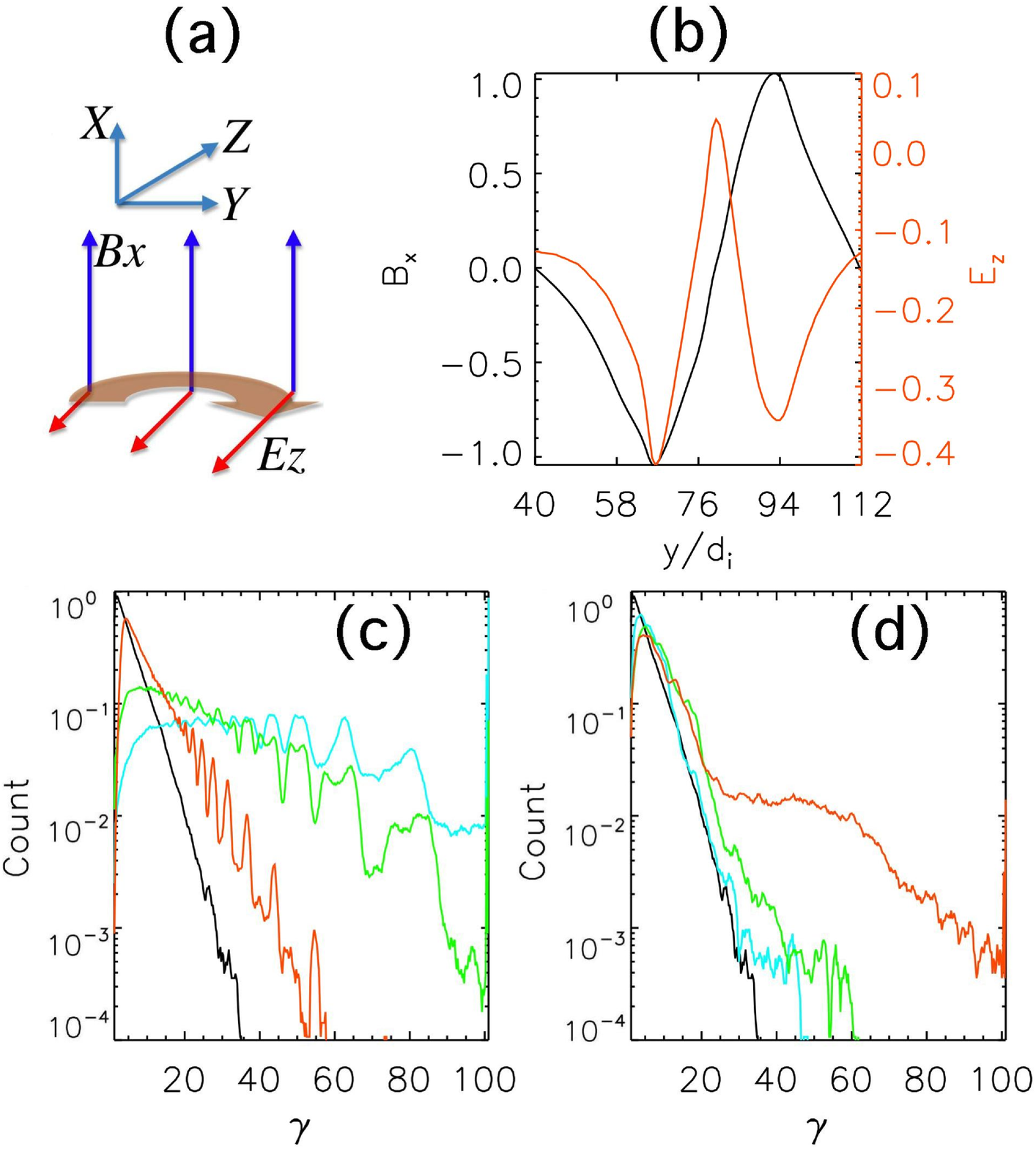}}
\caption{\label{fig:acceleration2} 
(Color)~(\emph{a}) Schematic view of the acceleration mechanism due to the variation of $E_z$ (constant $B_x$). (\emph{b}) The $y$ profile of $B_x$ (\emph{black}) and $E_z$ (\emph{red}) at $t\Omega_c=329$ of run A. (\emph{c}) Normalized particle distribution \emph{v.s.} energy $\gamma$ with different $B_x$ strength 
and a fixed $E_z$ profile: \emph{blue}, $B_x=0.7$; \emph{green}, $B_x=1.0$; \emph{red}, $B_x=3.0$; \emph{black}: normalized initial particle distribution for comparison. Larger $B_x$ leads to higher energy threshold (Eq.~\ref{eq1}), therefore fewer low-energy particles are accelerated. (\emph{d}) Particle distribution evolution \emph{v.s.} energy $\gamma$ using time-dependent field data from the simulation run A. \emph{black}, $t\Omega_c=0$; \emph{blue}, $t\Omega_c=300$; \emph{green}, $t\Omega_c=500$; \emph{red}, $t\Omega_c=750$. 
Significant non-thermal high-energy particles are obtained at $t\Omega_c\sim750$ (\emph{red} line). 
  }
\end{center}
\end{figure}

\begin{figure}[htbp]
\begin{center}
\scalebox{0.3}{\includegraphics{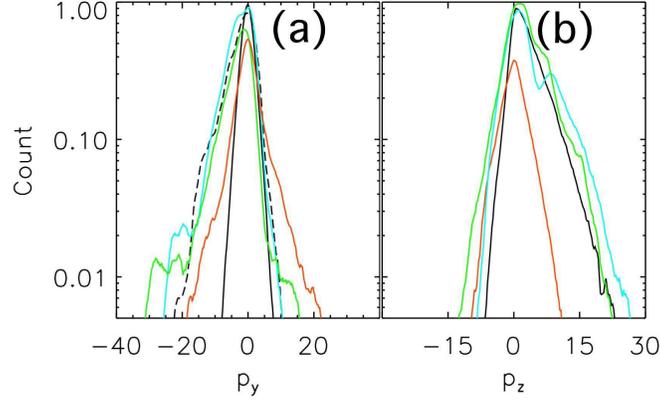}}
\caption{\label{fig:thermalization}(Color)~Particle number distribution \emph{v.s.} $p_y$ (\emph{a}) and electron number distribution  \emph{v.s.} $p_z$ (\emph{b}) of run A, taken from a box with $\Delta x=\Delta y=2d_i$  centered at $x/d_i=0$ and $y=89d_i$, at the right edge of the central island.  All data are normalized. (\emph{a}): \emph{black solid},~$t\Omega_c=0$;  \emph{black dash},~$t\Omega_c=295$; \emph{blue},~$t\Omega_c=311$; \emph{green},~$t\Omega_c=329$; \emph{red},~$t\Omega_c=348$. (\emph{b}): \emph{black solid},~$t\Omega_c=0$;  \emph{blue},~$t\Omega_c=329$; \emph{green},~$t\Omega_c=425$; \emph{red},~$t\Omega_c=1811$. 
}
\end{center}
\end{figure}

\begin{figure}[tbp]
\begin{center}
\scalebox{0.15}{\includegraphics{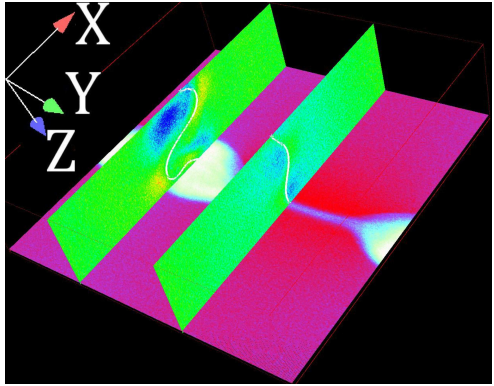}}
\caption{\label{fig:second_kink}(Color)~Spatial distribution of plasma density and fields at the nonlinear saturation stage from run C 
with $t\Omega_c\sim954$ . There are three cutting planes. One in the $\{x,y\}$ plane with contours of particle density $\rho\in[0.09,0.4]$ (from pink to white) at a $z$ location near the boundary, indicating the elongated reconnection site along $y$ between the large magnetic islands. The other two are in the $\{x,z\}$ plane with contours of reconnection magnetic field $B_y$ (white solid lines), colored by accelerating electrical field $E_z\in[-0.25,0.24]$ (from blue to yellow) at magnetic island ($y/d_i=60$) and at reconnection site ($y/d_i=134$) respectively. The run D with $L_z=200d_i$ gives the same conclusions about 
the second kink instability with similar wavelength. }
\end{center}
\end{figure}

\end{document}